\documentclass[journal]{IEEEtran}

\usepackage{amsmath}
\usepackage{amssymb}
\usepackage{graphicx}
\usepackage{epstopdf}
\usepackage{subfigure}
\usepackage{upgreek}
\usepackage{bm}
\usepackage{multirow}
\usepackage[ruled,lined,commentsnumbered,boxed]{algorithm2e}
\usepackage{url}
\usepackage{color}
\usepackage{soul}
\begin{document}
\title{Improving EEG Decoding via Clustering-based Multi-task Feature Learning}
\author{Yu Zhang*,
              Tao Zhou,
              Wei Wu,
              Hua Xie,
              Hongru Zhu,
              Guoxu Zhou,
              Andrzej Cichocki, \IEEEmembership{Fellow,~IEEE}
\thanks{G. Zhou was supported by the National Natural Science Foundation of China under Grant 61673124, Grant 61903095, and Guangdong Natural Science Foundation under Grant 2020A151501671. A. Cichocki was also supported by the Ministry of Education and Science of the Russian Federation (Grant 14.756.31.0001) and the Polish National Science Center (Grant 2016/20/W/N24/00354). \emph{*Corresponding author: Yu Zhang, email: yuzi20@lehigh.edu}}
\thanks{Y. Zhang is with the Department of Bioengineering, Lehigh University, Bethlehem, PA 18015, USA.}
\thanks{T. Zhou is with the Inception Institute of Artificial Intelligence, Abu Dhabi, UAE.}
\thanks{W. Wu is with School of Automation Science and Engineering, South China University of Technology, Guangzhou, China.}
\thanks{H. Xie is with Department of Psychiatry and Behavioral Sciences, Stanford University, Stanford, CA 94305, USA.}
\thanks{H. Zhu is with the Mental Health Center and Psychiatric Laboratory, the State Key Laboratory of Biotherapy, and also with Huaxi Brain Research Center, West China Hospital of Sichuan University, Chengdu 610041, China.}
\thanks{G. Zhou is with the School of Automation at Guangdong University of Technology, Guangzhou, China.}
\thanks{A. Cichocki is with the Skolkowo Institute of Science and Technology (SKOLTECH), Moscow, Russia, Tokyo University of  Agriculture and Technology (TUAT), Tokyo, Japan, and the Nicolaus Copernicus University (UMC), Torun, Poland.}
}

\markboth{IEEE Transactions on Neural Networks and Learning Systems, 2020}%
{Shell \MakeLowercase{\textit{et al.}}: Bare Demo of IEEEtran.cls for Journals}

\maketitle

\begin{abstract}
Accurate electroencephalogram (EEG) pattern decoding for specific mental tasks is one of the key steps for the development of brain-computer interface (BCI), which is quite challenging due to the considerably low signal-to-noise ratio of EEG collected at the brain scalp. Machine learning provides a promising technique to optimize EEG patterns toward better decoding accuracy. However, existing algorithms do not effectively explore the underlying data structure capturing the true EEG sample distribution, and hence can only yield a suboptimal decoding accuracy. To uncover the intrinsic distribution structure of EEG data, we propose a clustering-based multi-task feature learning algorithm for improved EEG pattern decoding. Specifically, we perform affinity propagation-based clustering to explore the subclasses (i.e., clusters) in each of the original classes, and then assign each subclass a unique label based on a one-versus-all encoding strategy. With the encoded label matrix, we devise a novel multi-task learning algorithm by exploiting the subclass relationship to jointly optimize the EEG pattern features from the uncovered subclasses. We then train a linear support vector machine with the optimized features for EEG pattern decoding. Extensive experimental studies are conducted on three EEG datasets to validate the effectiveness of our algorithm in comparison with other state-of-the-art approaches. The improved experimental results demonstrate the outstanding superiority of our algorithm, suggesting its prominent performance for EEG pattern decoding in BCI applications.
\end{abstract}

\begin{IEEEkeywords}
Brain-computer interface, Electroencephalogram, Multi-task learning, Motor imagery, Subclass regularization.
\end{IEEEkeywords}

\section{Introduction}
\IEEEPARstart{A}s an advanced communication technique, brain-computer interface (BCI) provides a promising approach for establishing a direct connection between a human brain and a computer \cite{mcfarland2017eeg} \cite{zhang2012novel}. The principal function of BCI is to translate the neural responses of a user into a computer command for external device operation, by decoding the brain activity patterns from a specific mental task \cite{wolpaw2000brain}. Through BCI, the user can gain an augmented communication ability that can be used to control the external environment, such as for wheelchair navigation, neuroprosthesis operation, and robot controlling \cite{huang2012electroencephalography,vidaurre2016eeg,ma2015novel}. In particular, BCIs have proven their efficacy as assistive and rehabilitative technologies bringing significant benefits and enhancing the quality of life for disabled patients \cite{chaudhary2016brain,ang2015randomized}.

One of the most commonly adopted brain activity patterns for BCI development is sensorimotor rhythm (SMR) \cite{pfurtscheller2006mu}, induced by performing a motor imagery (MI) task, such as imagining left or right hand movements. SMR can be measured using electroencephalogram (EEG) at a frequency band 8--13 Hz (or even a wider range) over a sensorimotor area of the brain \cite{pfurtscheller1997motor}. By decoding the SMR from EEG signals, an MI-based BCI can be developed to convert motor imagery tasks into computer commands for controlling an external device, which can serve as an assistive tool to strength control ability or help restore motor function \cite{pichiorri2015brain,mcfarland2008brain}. However, accurate decoding of SMR is quite challenging because of the fact that EEG signals generally present considerably low signal-to-noise ratio due to various noise disturbances \cite{Blankertz1,lotte2018review}. Thus, developing a sophisticated algorithm for SMR pattern optimization is currently one of the research hotspots in the field of BCI \cite{park2017filter,zhang2018multi,qiu2016improved}.

Machine learning techniques have been shown to provide promising multivariate analysis tools for the improvement of neural pattern decoding. In the past decade, an increasing number of methods have been designed for EEG pattern analysis in various BCI applications \cite{jin2018eeg,zhang2018temporally,jiao2018sparse,wu2019regularized,zhang2018two,li2016unified}. Common spatial pattern (CSP) is the most commonly used spatial filtering method to improve the signal-to-noise ratio of SMR \cite{Blankertz1}. CSP aims at estimating pairs of projection vectors (also known as spatial filters) from multi-channel EEG signals of two classes (e.g., two different MI tasks). With the derived projection vectors, the original signals are then transformed into a low dimensional feature space that maximizes the variance ratio of the transformed signals from one class versus another class. Since the most prominent property of SMR is the band-power change, the CSP-based spatial filtering provides an effective way to capture the potential the discriminant information of EEG pattern associated to MI tasks \cite{Blankertz1}.

So far, numerous research efforts have been dedicated to developing variants of CSP towards further improved EEG pattern decoding accuracy. By exploiting the mutual information between the extracted CSP features, a filter bank CSP (FBCSP) \cite{FBCSP} was designed to optimize CSP features from multiple filter bands. By incorporating Fisher's ratio, a discriminant extension of FBCSP (DFBCSP) \cite{DFBCSP} was further proposed to select the most discriminative filter bands, and hence enhance the separability of CSP features between classes. A sparsity-constrained filter band common spatial pattern (SFBCSP) \cite{SFBCSP} was also developed to automatically explore a compact combination of multi-band CSP features via a sparse learning strategy for improved pattern separability. Some other variants of CSP can be found in the literature \cite{samek2012stationary,park2014augmented,lotte2011regularizing,higashi2013simultaneous}.

Although the above-mentioned methods have shown their strengths improving the SMR decoding accuracy to varying degrees, none of them effectively explored the important data structure that captures the true EEG sample distribution. A model derived in such a way is likely degrade the learning performance since the underlying distribution structure of data is generally unknown in advance since EEG signals are highly nonstationary with large trial-to-trial variations \cite{arvaneh2013optimizing,Muller2008ML}. Intuitively, discovering the intrinsic subclasses in EEG data could provide more prior information for the model learning, enabling better neural pattern decoding. Incorporating subclass analysis into the feature optimization of magnetic resonance imaging has been proposed to mitigate the effects of heterogeneity and improve diagnosis performance for brain disorders \cite{suk2014subclass,liu2016inherent}. However, to the best of our knowledge, no previous methods have employed the potential distribution structure information of data in EEG pattern decoding, which might bring significant benefits and enable to achieve higher classification accuracy for improved performance in BCI applications.

To this end, we propose a clustering-based multi-task learning framework to effectively uncover and exploit the inherent distribution structure of the data, for more accurate MI-related EEG classification. Specifically, we first divide samples from each original class into multiple sub-classes (i.e, clusters) using afinity propagation (AP) clustering algorithm \cite{frey2007clustering}. Then, we encode each subclass with a unique label vector to transform the two-class problem into a multi-class problem. We further develop a novel multi-task learning algorithm by exploiting the subclass relationships to jointly select the most discriminative features from the uncovered subclasses. We name the developed framework as subclass relationship regularized multi-task learning (srMTL). Using three public EEG datasets, we carry out extensive experimental comparisons between our proposed algorithm and other state-of-the-art methods. The superior classification accuracy of our algorithm demonstrates its promise in the development of advanced BCI systems.

The main contributions of our study can be summarized as follows: (1) We first propose to exploit AP clustering for exploring the underlying structure of EEG data; (2) A clustering-based multi-task learning algorithm is developed for feature optimization based on the uncovered subclasses; (3) A subclass relationship regularization is further incorporated into the learning model to capture the manifold structure of subclasses for improved EEG pattern decoding. The rest of the paper is organized as follows. In Section II Methodology, we explain in detail the proposed srMTL algorithm, including the subclass generation and multi-task learning model, preceded by briefly introducing the feature extraction of EEG pattern. In Section III Experimental Study, we describe the EEG data acquisition, performance evaluation and comparative results. In Section IV Discussion, we assess the advantages of srMTL based on the experimental results obtained, analyze the parameter sensitivity, investigate the computational efficiency, and discuss the potential extensions of our algorithm for future studies. Finally, we summarize our study in Section V Conclusions.

\begin{figure*}[!t]
	\centering
	\includegraphics[width=0.8\textwidth]{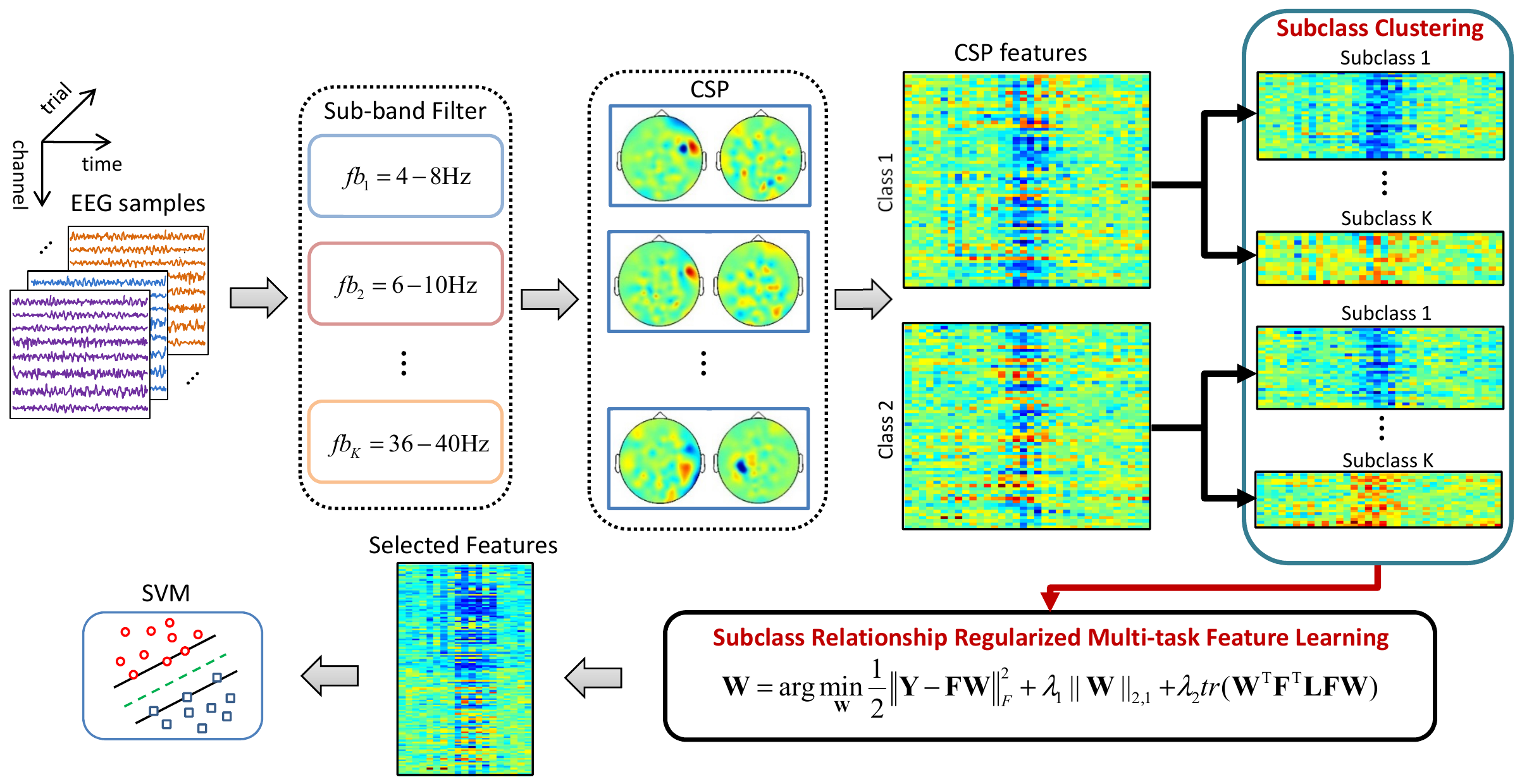}
	\caption{Overall framework of the proposed srMTL method for MI-related EEG classification.}
	\label{F-Framework}
\end{figure*}

\section{Methodology}
\subsection{CSP feature extraction of SMR}
Due to the effects of volume conduction and noise disturbance, EEG signals collected at the brain scalp generally present considerably low signal-to-noise ratio. Simply extracting features from a single or bipolar channel at the sensorimotor area typically yields a relatively poor decoding accuracy of SMR \cite{Blankertz1}. To improve the feature discriminability, spatial filtering has been proposed for the multi-channel optimization of EEG signals. Specifically, CSP has been shown to provide an effective approach for extracting the discriminative feature from an SMR pattern by maximizing the difference between variance of the spatial filtered signals from different classes \cite{ramoser2000optimal}.

Assume $\mathbf{X}_{i,l} \in \mathbb{R}^{C \times P}$ consists of EEG signals collected from trial $i$ of class $l$, with $C$ channels and $P$ temporal points. After bandpass filtering at a specific frequency band, CSP implements spatial filter optimization by solving
\begin{equation}\label{E-CSP}
\mathop{\max}\limits_{{\bf u}}~J({\bf u}) = \frac{{\bf u}^T {\bf \Sigma}_1{\bf u}}{{\bf u}^T {\bf \Sigma}_2{\bf u}} ~~~~\mbox{s.t.}~~\| {\bf u} \|_2=1,
\end{equation}
where ${\bf u} \in \mathbb{R}^C$ is a spatial filter, $\| \cdot \|_2$ is the $l_2$-norm, and $\mathbf{\Sigma}_l$ denotes the spatial covariance matrix of class $l$ and is computed as
\begin{equation}
{\bf \Sigma}_l = \frac{1}{N_l}\sum\limits_{i=1}^{N_l}{\bf X}_{i,l}{\bf X}_{i,l}^T,
\end{equation}
with $N_l$ being the number of samples in class $l$. The optimization (\ref{E-CSP}) can be achieved by equivalently solving a generalized eigenvalue problem: ${\bf \Sigma}_1 \mathbf{u} = \lambda {\bf \Sigma}_2 \mathbf{u}$. A set of spatial filters $\mathbf{U}=[\mathbf{u}_1,\ldots,\mathbf{u}_{2M}]$ is then derived by collecting eigenvectors corresponding to the $M$ largest and smallest generalized eigenvalues. For a given EEG sample $\mathbf{X} \in \mathbb{R}^{C \times P}$, we can extract a feature vector $\mathbf{f}=[f_1,\ldots,f_{2M}]^T$ with entries:
\begin{equation} \label{E-Feat}
f_m=\log(\mbox{var}(\mathbf{u}_m^T \mathbf{X})),~~m=1,\ldots,2M,
\end{equation}
where $\mbox{var}(\cdot)$ denotes the variance.

Instead of using a single frequency band for bandpass filtering, multi-band optimization methods have been recommended to extract more discriminative CSP features, and hence provide better accuracy of SMR classification \cite{FBCSP,DFBCSP,SFBCSP}. Suppose EEG data is bandpass filtered at $G$ filter bands, an augmented feature vector can be formed as:
\begin{equation}\label{E-AugFeat}
\mathbf{\hat f}=[f_1,\ldots,f_{2M},f_{2M+1},\ldots,f_{2MG}]^T
\end{equation}
by concatenating the features extracted from each filter band. With an appropriately designed feature selection method, such as the mutual information based best individual feature selection algorithm \cite{FBCSP} or sparse learning algorithm \cite{SFBCSP}, we can optimize the extracted CSP features by selecting the ones with relatively high discriminative information for the subsequent SMR decoding. A typical multi-band optimization method, SFBCSP \cite{SFBCSP}, has been designed to select informative features through sparse regression \cite{tibshirani1996regression}:
\begin{equation} \label{E-Sparse}
\mathbf{w}=\arg\min_{\mathbf{w}} \frac{1}{2}\| \mathbf{y}-\mathbf{F}\mathbf{w} \|_2^2 + \lambda\| \mathbf{w} \|_1,
\end{equation}
where $\mathbf{y} \in \mathbb{R}^N$ denotes a label vector containing class labels $-1$ or $1$, each row of the feature matrix $\mathbf{F} \in \mathbb{R}^{N \times D}$ ($D=2MG$) consists of the augmented feature vector extracted according to (\ref{E-AugFeat}), and the $l_1$-norm $\| \mathbf{w} \|_1$ penalizes the weight vector $\mathbf{w} \in \mathbb{R}^D$ to be sparse so that the weights corresponding to non-informative features are restricted to be zeros. As a result, SFBCSP prunes CSP features to best represent the label space.

\subsection{Subclass exploration via clustering}
Although the multi-band filter bank methods have improved the classification accuracy to some extent, most of them simply ignored the intrinsic structure of EEG data. Without uncovering the true sample distribution, which is generally not pre-known, the derived feature optimization model can only yield a suboptimal result. For example, as shown in Fig. \ref{F-Framework}, a high between-trial variability can be observed in the CSP feature set. The feature patterns of some trials are even more similar to those from different classes than those from the same class. Thus, simply implementing feature selection in a binary-class way may not provide the optimal decoding accuracy.

In order to exploit the intrinsic structure of EEG data for more accurate decoding of neural patterns, we apply AP clustering \cite{frey2007clustering}, which explores the potential subclasses in each of the original classes. The distinct clusters obtained indicate various intrinsic sample distributions hidden in the original classes. By treating each of the clusters as a subclass, we then encode each subclass with an unique label based on the one-versus-all encoding strategy to form a label matrix $\mathbf{Y}$. Assuming we discover $K$ subclasses, the label matrix is encoded as $\mathbf{Y}=[\mathbf{y}_1,\mathbf{y}_2,\ldots,\mathbf{y}_N]^T \in \mathbb{R}^{N \times K}$, where the $k$-th element in $\mathbf{y}_i$ is defined as:
\begin{equation} \label{E-LabelMat}
y_{ik}=\left\{ {\begin{array}{*{20}{l}}
1, ~{\rm if}~{\rm sample}~\mathbf{x}_i \in {\rm cluster}~k \\
0, ~{\rm otherwise}
\end{array}} \right.
\end{equation}

\begin{figure}[!t]
	\centering
	\includegraphics[width=0.45\textwidth]{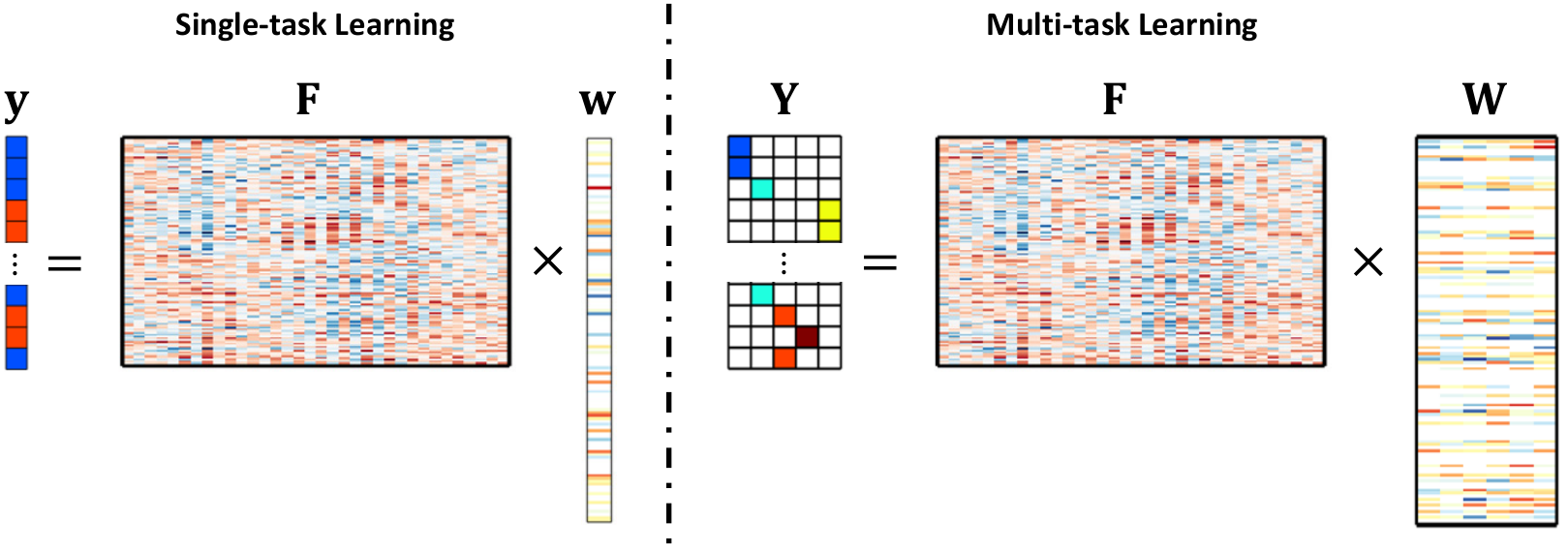}
	\caption{Illustration of single-task learning versus multi-task learning. The single-task learning estimates a single weight vector based on a single target vector. In contrast, the multi-task learning estimates a weight matrix based on an encoded target matrix, which allows us to optimize features jointly across multiple tasks.}
	\label{F-MTL}
\end{figure}

\subsection{Subclass relationship regularized multi-task learning}
The sparse feature selection formulated in (\ref{E-Sparse}) is typically a single-task learning model (as shown in Fig. \ref{F-MTL}(1)), which estimates a single weight vector based on a single target vector, and hence is not able to address the regression problem on multiple targets when taking the subclass information into account. To achieve a joint feature selection across all the uncovered subclasses, we consider exploiting the multi-task learning (MTL) strategy \cite{liu2009multi} (as shown in Fig. \ref{F-MTL}(2)). To this end, we devise a novel algorithm based on MTL with the subclass label matrix $\mathbf{Y}$ encoded using (\ref{E-LabelMat}). First, we formulate a subclass-based MTL model as a $l_{2,1}$-norm penalized sparse regression:
\begin{equation} \label{E-MTL}
\mathbf{W}=\arg\min_{\mathbf{W}} \frac{1}{2}\| \mathbf{Y}-\mathbf{F}\mathbf{W} \|_F^2 + \lambda\| \mathbf{W} \|_{2,1},
\end{equation}
where $\| \cdot \|_F$ denotes the Frobenius norm, $\mathbf{W}\in\mathbb{R}^{D \times K}$ is a weight matrix, and the $l_{2,1}$-norm $\| \mathbf{W} \|_{2,1} = \sum\nolimits_i\sqrt{\sum\nolimits_j w_{ij}^2}$ forces some rows of $\mathbf{W}$ to be zeros, deriving a joint feature selection for multiple subclasses. The hyperparameter $\lambda$ controls the group sparsity of $\mathbf{W}$. With the subclass label information, the MTL model allows us to select features that are separable across all subclasses. Beside the subclass label information, we also consider that the intrinsic subclass relationship that characterizes the manifold structure of subclasses may also be helpful to improve the decoding of neural patterns. For instance, if two samples $\mathbf{f}_i$ and $\mathbf{f}_j$ belong to the same cluster, they should also be close to each other after being projected to the encoded label space. To preserve the potential subclass relationship, we introduce the following graph Laplacian regularization:
\begin{equation}
\Omega=\sum_{i,j}^N s_{ij} \| \mathbf{W}^T\mathbf{f}_i - \mathbf{W}^T \mathbf{f}_j \|_2^2 = tr(\mathbf{W}^T \mathbf{F}^T\mathbf{L}\mathbf{F}\mathbf{W}),
\end{equation}
where $tr(\cdot)$ denotes the trace, $\mathbf{L}=\mathbf{D}-\mathbf{S}$, $\mathbf{S}=[s_{ij}] \in \mathbb{R}^{N \times N}$ is a similarity matrix characterizing the relationship between subclasses, $\mathbf{D} \in \mathbb{R}^{N \times N}$ is a diagonal matrix with its diagonal entries defined as $d_{ii}=\sum\nolimits_j s_{ij}$, and $s_{ij}$ is defined as:
\begin{equation}
s_{ij}=\left\{ {\begin{array}{*{20}{l}}
1, ~{\rm if}~\mathbf{f}_i~{\rm and}~\mathbf{f}_j~{\rm are~from~the~same~cluster} \\
0, ~{\rm otherwise}
\end{array}} \right.
\end{equation}

By incorporating the graph Laplacian regularization $\Omega$ as a penalty term into the MTL model, we propose a subclass relationship regularized MTL (srMTL) algorithm formulated as the following optimization problem:
\begin{align} \label{E-srMTL}
\mathbf{W} = & \arg\min_{\mathbf{W}} \frac{1}{2}\| \mathbf{Y}-\mathbf{F}\mathbf{W} \|_F^2 + \lambda_1 \| \mathbf{W} \|_{2,1} \nonumber \\
& ~~~~~~~~~~~~ + \lambda_2 tr(\mathbf{W}^T \mathbf{F}^T \mathbf{L}\mathbf{F}\mathbf{W}),
\end{align}
where $\lambda_2$ is the hyperparameter that controls the effect of the subclass relationship on the loss function. Unlike the single-task learning, the proposed srMTL algorithm formulates the feature optimization as a multi-class problem for joint feature selection by exploiting both the label and structure information of subclasses, which may help further enhance the SMR decoding accuracy.

With the $l_{2,1}$-norm regularization, some rows of the optimal solution $\mathbf{W}$ will be forced to be zeros, indicating the corresponding features are not informative for classifying the subclasses. Accordingly, we only retain those features whose weight vectors are non-zero for the subsequent classification analysis. A linear SVM classifier is trained on the selected features and applied to the decoding of the SMR pattern.

\begin{algorithm}[!t] \label{A-srMTL}
\KwIn{$\mathbf{L} \in \mathbb{R}^{N\times N}$, the step size $\mu=1$, the coefficient $\alpha(0)=1$, the maximum iteration number $T$, and the hyperparameters $\lambda_1$ and $\lambda_2$.}
\KwOut{Learned weight matrix $\mathbf{W}(t+1) \in \mathbb{R}^{D \times C}$}
    \quad \\
   Initialize $\mathbf{W}(0)=\mathbf{W}(1)=\mathbf{1}_{D\times T}$; \\
   \For{$t=1$ to $T$}{
        $\alpha(t)=\frac{1+\sqrt{1+4\alpha(t-1)^2}}{2}$;
        $\mathbf{P}(t)=\mathbf{W}(t)+\frac{\alpha(t-1)-1}{\alpha(t)}(\mathbf{W}(t)-\mathbf{W}(t-1))$; \\
        \For{$d=1$ to $D$}{
            $\mathbf{v}^d(t)=\mathbf{p}^d(t)-\frac{1}{\mu}\nabla f(\mathbf{p}^d(t))$;  \\
            $\mathbf{w}^d(t+1)=\arg\min\limits_{\mathbf{w}^d}\frac{1}{2}\| \mathbf{w}^d-\mathbf{v}^d(t) \|_2^2 + \frac{\lambda_1}{\mu}\| \mathbf{w}^d \|_2$; \\
        }
        Find the smallest $\mu=\mu,2\mu,\ldots$ such that\\
         $f(\mathbf{W}(t+1))+\mathcal{L}(\mathbf{W}(t+1)) < \mathcal{G}_{\mu}(\mathbf{W},\mathbf{P}(t))$; \\
    }
\caption{Optimization algorithm for the proposed srMTL algorithm}
\end{algorithm}

\subsection{Optimization}
Although the objective function in (\ref{E-srMTL}) is convex, it is challenging to solve because of the non-smooth penalty term $\|\mathbf{X}\|_{2,1}$. In this study, we adopt the accelerated proximal gradient (APG) \cite{parikh2014proximal} method to solve this optimization problem. To this end, we separate the objective function into a smooth part:
\begin{equation}
f(\mathbf{W})=\frac{1}{2} \| \mathbf{Y}-\mathbf{F}\mathbf{W} \|_F^2 + \lambda_2tr(\mathbf{W}^T\mathbf{F}^T\mathbf{L}\mathbf{F}\mathbf{W}),
\end{equation}
and a non-smooth part:
\begin{equation}
\mathcal{L}(\mathbf{W})=\lambda_1\| \mathbf{W} \|_{2,1},
\end{equation}
To approximate the composite function $f(\mathbf{W})+\mathcal{L}(\mathbf{W})$, we construct the following function:
\begin{align}
\mathcal{G}_{\mu}(\mathbf{W},\mathbf{W}(t)) = & f(\mathbf{W}(t)) + \langle\mathbf{W}-\mathbf{W}(t),\nabla f(\mathbf{W}(t))\rangle \nonumber \\
& + \frac{\mu}{2}\| \mathbf{W}-\mathbf{W}(t) \|_F^2 + \mathcal{L}(\mathbf{W}),
\end{align}
where $\nabla f(\mathbf{W}(t))$ is the gradient of $f(\mathbf{W})$ at the point $\mathbf{W}(t)$, $\langle \cdot,\cdot \rangle$ denotes an inner product operator, and $\mu$ is a step size that can be determined by the line search. The iterative update rule of the proximal gradient method is then given by:
\begin{equation}
\mathbf{W}(t+1)=\arg\min_{\mathbf{W}} \frac{1}{2} \| \mathbf{W}-\mathbf{V}(t) \|_F^2 + \frac{1}{\mu}\mathcal{L}(\mathbf{W}),
\end{equation}
where $\mathbf{V}(t)=\mathbf{W}(t)-\frac{1}{\mu}\nabla f(\mathbf{W}(t))$ and $\mathbf{W}(t)$ is the search point of $\mathbf{W}$ obtained at the $k$-th iteration. Since each row of $\mathbf{W}$ is decoupled, we can update the weights for each row individually:
\begin{equation}\label{E-PO}
\mathbf{w}^d(t+1) = \arg\min_{\mathbf{w}^d}\frac{1}{2}\| \mathbf{w}^d-\mathbf{v}^d(t) \|_2^2 + \frac{\lambda_1}{\mu}\|\mathbf{w}^d\|_2,
\end{equation}
where $\mathbf{v}^d(t)$ denotes the $d$-th row of $\mathbf{V}(t)$. We resort to a closed form solution from the literature and estimate $\mathbf{w}^d(t+1)$ as:
\begin{equation}
{{\mathbf{\tilde w}}^d} = \left\{ {\begin{array}{*{20}{c}}
\left( {1 - \frac{{{\lambda _1}}}{{\mu\|\mathbf{v}^d(t)\|_2^2}}} \right)\mathbf{v}^d(t), & {\rm{if}}~\|\mathbf{v}^d(t)\|_2^2 > \frac{\lambda _1}{\mu}\\
0,&\rm{otherwise},
\end{array}} \right.
\end{equation}

\begin{figure}[!t]
	\centering
	\includegraphics[width=0.45\textwidth]{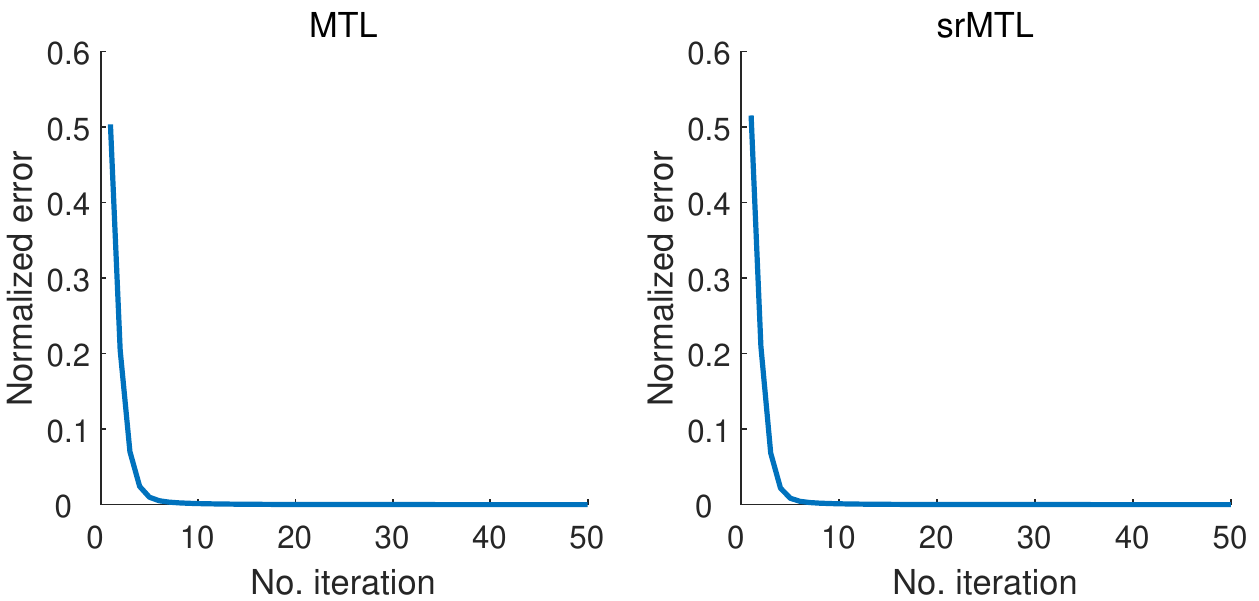}
	\caption{Convergence curves of MTL and srMTL algorithms. The normalized error is defined as $\|f(t)-f(t+1)\|_2/f(t)$ where $f(t)$ denotes the objective function at $t$th iteration.}
	\label{F-Convergence}
\end{figure}

To accelerate the proximal gradient method, we replace the search point $\mathbf{W}(t)$ by:
\begin{equation}
\mathbf{P}(t)=\mathbf{W}(t)+\frac{\alpha(t-1)-1}{\alpha(t)}(\mathbf{W}(t)-\mathbf{W}(t-1)),
\end{equation}
where $\alpha(t)$ is a properly selected coefficient and usually set as $\alpha(t)=\frac{1+\sqrt{1+4\alpha(t-1)^2}}{2}$. Algorithm \ref{A-srMTL} summarizes the optimization procedure of the proposed algorithm. With the APG-based iterative algorithm, the optimization can be achieved with convergence rate of $\mathcal{O}(\frac{1}{t^2})$ for a fix iteration number $t$. To validate the convergence of our algorithm, we provide an example (see Fig. \ref{F-Convergence}) to depict how the iterative error of the objective function change with respective to the increase in the number of iteration.  For both MTL and srMTL algorithms, the iterative error decrease rapidly and achieve convergence within 50 iterations.

\begin{table*}[!t]
\renewcommand{\arraystretch}{1.2}
\scriptsize
\caption{Comparison of EEG pattern decoding accuracy (\%) between different algorithms on the BCI Competition IV Dataset IIa. A $5 \times 5$-fold cross-validation is performed to evaluate the performance. Boldface indicates the highest accuracy for each of the subjects. The $p$ values are obtained by implementing paired sample t-test with false discovery rate correction between our algorithm and each of the other methods.} \vspace*{-12pt}\label{T-Acc1}
\begin{center}
\def\temptablewidth{0.6\textwidth}
{\rule{\temptablewidth}{1pt}}
\begin{tabular*}{\temptablewidth}{@{\extracolsep{\fill}}ccccccc}
Subject     &     CSP   &     FBCSP   &   DFBCSP    &     SFBCSP     &     MTL      &    srMTL              \\
\hline
B0103T    &   76.6    &   77.3    &   76.0        &    78.1       &    79.5     &       \textbf{80.1}            \\
B0203T    &   56.4    &   54.8    &   55.4        &    57.1       &   \textbf{60.6}     &       \textbf{60.6}            \\
B0303T    &   51.3    &   52.0    &   53.2        &    53.8     &    53.8    &       \textbf{55.1}            \\
B0403T    &   98.6    &   98.6    &   98.6        &    98.8       &    98.9    &       \textbf{99.8}             \\
B0503T    &   83.5   &   88.1     &   91.8       &    90.8       &    91.3    &       \textbf{92.5}            \\
B0603T    &  67.0    &   80.1     &   81.1       &    81.8       &    81.8    &       \textbf{82.9}            \\
B0703T    &  83.9    &   89.7     &   90.1       &    91.4       &    92.8    &       \textbf{94.1}            \\
B0803T    &  86.9    &   89.0     &   89.1       &    89.3      &    90.3    &       \textbf{91.5}            \\
B0903T    &  81.9    &   84.3     &  85.5       &    85.8        &    87.5    &       \textbf{88.7}            \\
Average   &  76.2$\pm$15.2    &   79.3$\pm$15.9   &    80.1$\pm$16.0     &    80.8$\pm$15.5      &    81.3$\pm$15.2    &       \textbf{82.8}$\pm$15.4          \\
\hline
$p$-value   &  $p<0.01$   &   $p<0.01$    &    $p<0.01$     &    $p<0.01$     &    $p<0.01$     &    --      \\
\hline
\end{tabular*}
{\rule{\temptablewidth}{1pt}}
\end{center}
\end{table*}

\begin{table*}[!t]
\renewcommand{\arraystretch}{1.2}
\scriptsize
\caption{Comparison of EEG pattern decoding accuracy (\%) between different algorithms on the BCI Competition IV dataset 1. A $5 \times 5$-fold cross-validation is performed to evaluate the performance. Boldface indicates the highest accuracy for each of the subjects.} \vspace*{-12pt}\label{T-Acc2}
\begin{center}
\def\temptablewidth{0.6\textwidth}
{\rule{\temptablewidth}{1pt}}
\begin{tabular*}{\temptablewidth}{@{\extracolsep{\fill}}ccccccc}
Subject     &   CSP   &     FBCSP    &    DFBCSP  &       SFBCSP     &     MTL      &     srMTL              \\
\hline
a      &   81.8   &     82.0   &   83.2  &    85.0     &    86.8    &      \textbf{89.2}            \\
b      &   57.2   &     59.8   &  61.1  &    62.2     &    62.5    &       \textbf{65.0}            \\
f       &   86.5   &     91.5   &  89.3  &    92.3     &    94.0    &       \textbf{94.7}           \\
g      &   90.1   &    91.8    &  92.0  &    92.1    &    93.2    &        \textbf{94.7}            \\
Average  &   78.9$\pm$14.9  &   81.3$\pm$15.0  &  81.4$\pm$14.0   &   82.9$\pm$14.2   &   84.1$\pm$14.8   &    \textbf{85.9}$\pm$14.2            \\
\hline
\end{tabular*}
{\rule{\temptablewidth}{1pt}}
\end{center}
\end{table*}

\begin{table*}[!t]
\renewcommand{\arraystretch}{1.2}
\scriptsize
\caption{Comparison of EEG pattern decoding accuracy (\%) between different algorithms on the BCI Competition III dataset IIIa. A $5 \times 5$-fold cross-validation is performed to evaluate the performance. Boldface indicates the highest accuracy for each of the subjects.} \vspace*{-12pt}\label{T-Acc3}
\begin{center}
\def\temptablewidth{0.6\textwidth}
{\rule{\temptablewidth}{1pt}}
\begin{tabular*}{\temptablewidth}{@{\extracolsep{\fill}}ccccccc}
Subject    &   CSP    &    FBCSP        &    DFBCSP      &     SFBCSP         &     MTL      &    srMTL              \\
\hline
K3      &  96.1   &    96.8     &  97.0    &    97.6       &     98.1      &     \textbf{99.2}            \\
K6      &   58.0     &   58.5        &  62.8       &    60.7       &      65.2     &      \textbf{67.8}           \\
L1      &   92.8      &    93.0        &    94.0      &    94.8       &     95.2     &      \textbf{96.3}           \\
Average  &   82.3$\pm$21.1      &    82.8$\pm$21.1   &   84.6$\pm$18.9     &    84.4$\pm$20.5      &       86.2$\pm$18.2    &       \textbf{87.8}$\pm$17.4            \\
\hline
\end{tabular*}
{\rule{\temptablewidth}{1pt}}
\end{center}
\end{table*}

\section{Experimental study}
\subsection{EEG data description}
\subsubsection{Dataset-1}
The first dataset used for our experimental study was from the BCI Competition IV dataset IIb. EEG data was collected from nine subjects at three electrodes C3, Cz and C4 with a sampling rate of 250 Hz, during right hand and left hand MI tasks. This study only used the third training sessions of the dataset, i.e., ``B0103T", ``B0203T", $\ldots$, ``B0903T". For each subject, a total of 160 trials of EEG measurements are available (half for each class of MI). In each trial, the subject was indicated by a visual cue to perform MI task for 4.5 s. See website \url{http://www.bbci.de/competition/iv/} for more details about the dataset.

\subsubsection{Dataset-2}
We used BCI Competition IV dataset 1 provided by the Berlin BCI group as the second dataset for testing our algorithm. EEG signals were recorded from four subjects ``a", ``b", ``f", and ``g" using BrainAmp MR plus amplifier and a Ag/AgCl cap with 59 electrodes, at a sampling rate of 1000 Hz with bandpass filtering between 0.05 and 200 Hz. Two classes of motor imagery were selected from the three classes (left hand, right hand, and foot) to instruct the subjects to perform the cued mental task for 4 s. After the whole experimental session, each subject completed 200 trials (half per class). The EEG signals were then downsampled to 100 Hz for the subsequent analysis. More details about the dataset can be found at \url{http://www.bbci.de/competition/iv/}.

\subsubsection{Dataset-3}
The third dataset was from the BCI Competition III dataset IIIa. EEG signals were recorded from three subjects, named ``K3", ``K6" and ``L1" using a Neuroscan amplifier at 60 electrodes with sampling rate of 250 Hz and bandpass filtering between 1 and 50 Hz. Line noise was suppressed by a notch filter. In the experiment, each subject was indicated to perform a sequential repetition of cue-based trials. In each trial, an arrow cue was presented to instruct the subject to imagine either a left hand, right hand, foot, or tongue movement. Each movement imagination task was performed for 4 s. Only the trials corresponding to left or right hand MI tasks were used for our study. The number of trials are 180, 120 and 120 for subjects ``K3", ``K6" and ``L1", respectively, and half of the trials were recorded during left hand MI tasks. More details about this dataset can be found at \url{http://www.bbci.de/competition/iii/}.

\subsection{Experimental evaluation and results}
To validate the effectiveness of the proposed srMTL algorithm, we implemented extensive experimental comparisons between our method and other state-of-the-art algorithms:
\begin{enumerate}
  \item CSP \cite{Blankertz1}: A baseline method that extracts CSP features at a single frequency band 4--40 Hz.
  \item FBCSP \cite{FBCSP}: Optimizes the multi-band CSP features using the mutual information best individual feature selection algorithm.
  \item DFBCSP \cite{DFBCSP}: Optimizes the multi-band CSP features using the Fisher's discriminant criteria.
  \item SFBCSP \cite{SFBCSP}: Optimizes the multi-band CSP features using the single-task learning based sparse regression model.
  \item MTL \cite{liu2009multi}: Optimizes the multi-band CSP features using the subclass-based multi-task learning algorithm.
  \item srMTL: Optimizes the multi-band CSP features using the subclass relationship regularized multi-task learning algorithm.
\end{enumerate}

For multi-band CSP feature extraction, we performed bandpass filtering on the raw EEG data using a set of overlapping subbands filters. A total of 17 subbands were chosen from the frequency range 4--40 Hz with a bandwidth of 4 Hz and overlapping rate of 2 Hz: 4--8 Hz, 6--10 Hz, \ldots, 36--40 Hz. CSP features were then extracted from each of the subbands and concatenated to form a composite feature vector for the subsequent feature optimization and classification. A $5\times5$-fold cross-validation was implemented to evaluate the classification accuracy. In each run of the cross-validation, an inner-loop cross-validation was performed to determine the optimal hyperparameters $\lambda_1$ and $\lambda_2$ $\in \{ 0.01, 0.05, 0.1, 0.5, 1, 5, 10, 15, 20, 25, 30, 35, 40, 45, 50, 55, 60 \}$. The soft-margin parameter $C$ of SVM was set as its default (i.e., $C=1$). Paired sample t-test with false discovery rate correction was adopted to investigate the statistical significance of the comparison.

Table \ref{T-Acc1} summarizes the classification accuracies obtained by the compared algorithms on the dataset-1. Compared with other single-task learning approaches, multi-task learning algorithms (i.e., MTL and srMTL) achieved improved performance. By incorporating the subclass relationship as a prior into the MTL model, our proposed srMTL algorithm further enhanced the classification accuracy and significantly outperformed all other competing methods ($p<0.01$, corrected with false discovery rate). Tables \ref{T-Acc2} and \ref{T-Acc3} summarize the experimental results evaluated on the dataset-2 and dataset-3, respectively. Consistent with the results in Table \ref{T-Acc1}, the proposed srMTL algorithm yielded the highest classification accuracy for all subjects, demonstrating its superiority in EEG pattern recognition, compared to other state-of-the-art methods.

\begin{figure*}[!t]
	\centering
	\includegraphics[width=\textwidth]{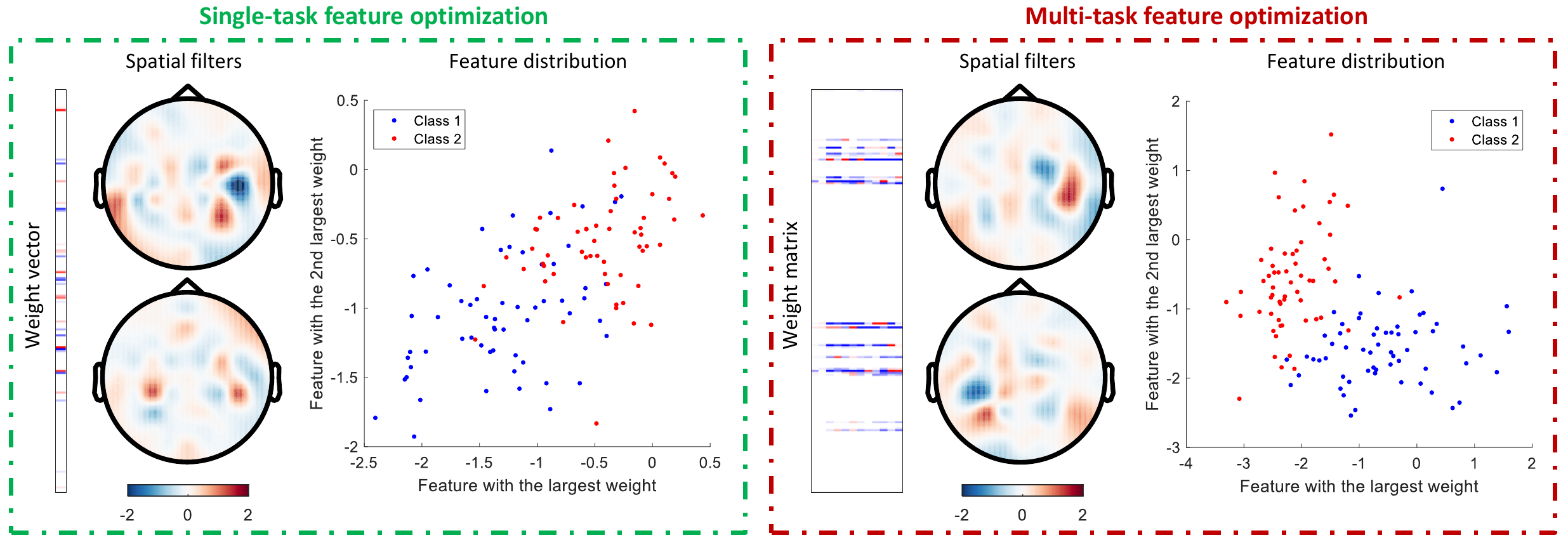}
	\caption{Comparison of sparse weights, spatial filers, and feature distribution between multi-task feature optimization (our method) and single-task feature optimization (SFBCSP). Unlike single-task learning, multi-task learning formulates the feature optimization as a multi-class problem for joint feature selection across the uncovered subclasses, and hence more accurately captures the most discriminative features around the sensorimotor areas for better SMR-related EEG pattern decoding.}
	\label{F-WeightSpatialFeat}
\end{figure*}

\section{Discussion}
In the past decades, numerous machine learning algorithms have been developed for EEG feature optimization and classification \cite{lotte2018review,zhang2019hierarchical,STDA,zhou2016linked,zhang2014frequency,chen2018novel,duan2020topological}. Most of these algorithms simply ignored the potentially important data structure that captures the true EEG sample distribution. As a result, a model derived in such a way may only provide a suboptimal performance of EEG decoding, since the underlying distribution structure of data is not pre-known due to the large trial-to-trial variations in the highly nonstationary EEG signals (as shown in Fig. \ref{F-Framework}) \cite{arvaneh2013optimizing,Muller2008ML}. To address this issue, our study exploited the idea of subclass analysis, which has been shown effective in addressing the heterogeneity issue in brain disease diagnosis \cite{suk2014subclass,liu2016inherent}. We incorporated the subclass analysis into multi-task learning and extended the SFBCSP (a typical single-task learning method) to achieve joint feature optimization across the uncovered subclasses that characterize the inherent distribution structure of the data. Unlike the single-task learning, multi-task learning allows us to jointly select the most significant features from multiple tasks and has demonstrated its strength in various applications \cite{zhu2016subspace,zhou2018Multiview,wang2017multi,zhang2018strength,zhu2017novel,zhu2016block,zhou2019multiview,zhou2019tmi,wang2018sparse} (also see the comparison in Fig. \ref{F-MTL}). By further taking the subclass relationships into account, our proposed srMTL algorithm achieved the highest performance for SMR decoding in comparison with other state-of-the-art approaches.

\subsection{Spatial pattern and feature distribution}
\begin{figure}[!t]
	\centering
	\includegraphics[width=0.4\textwidth]{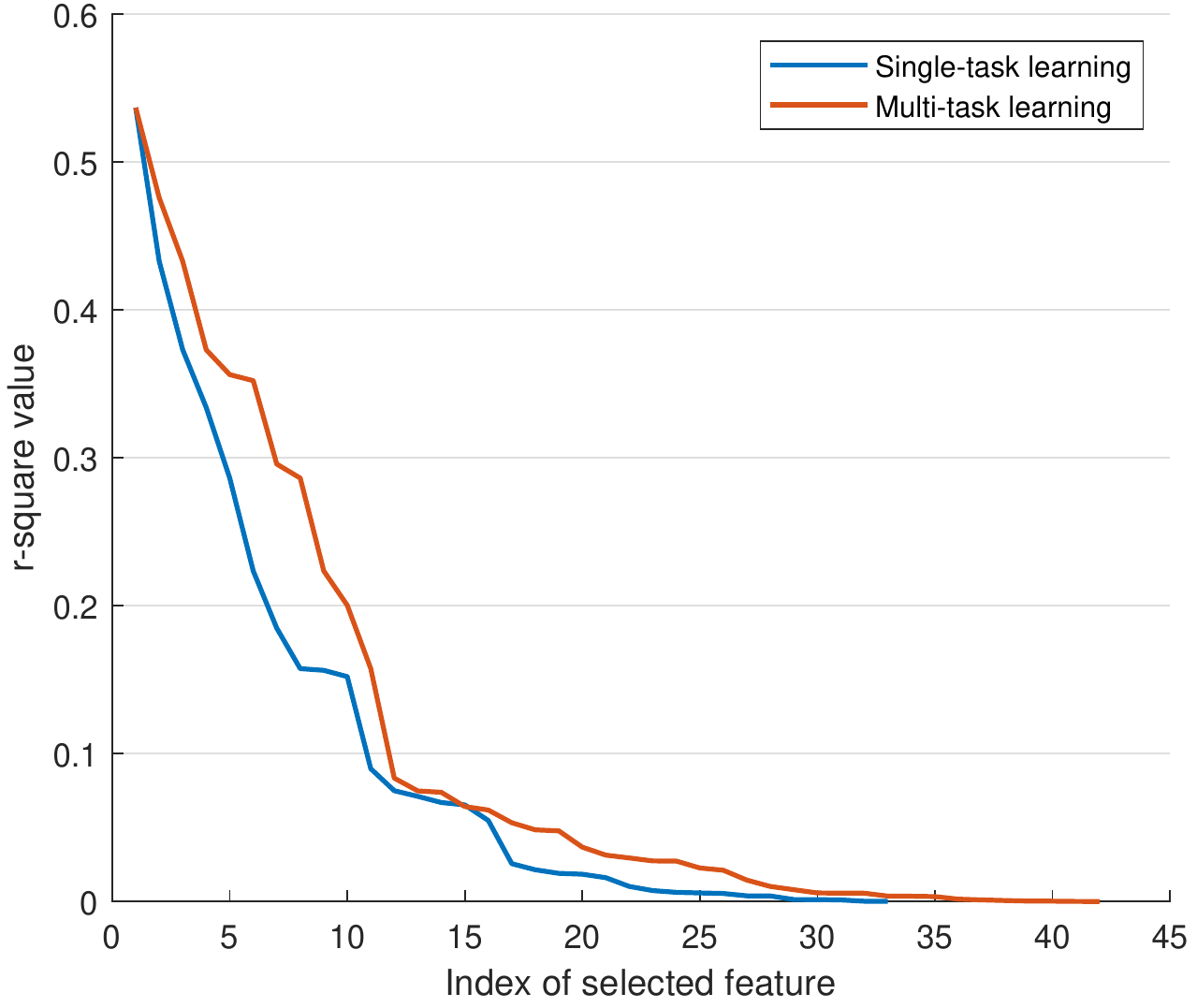}
	\caption{Comparison of feature discriminability. For each of the selected features, the discriminability is measured by r-square discriminant index. This indicates that multi-task learning (our method) selects more features with higher discriminability in comparison with single-task learning, which are conducive to better decoding accuracy.}
	\label{F-SelectedFeatures}
\end{figure}

To better interpret the experimental results, we visualized the regression weights, spatial filters, and the corresponding feature distributions derived by single-task feature optimization (i.e., SFBCSP) and multi-task feature optimization (our method), respectively (see Fig. \ref{F-WeightSpatialFeat}). Compared with single-task learning, our method more accurately captured the most significant features around the sensorimotor areas, achieving higher separability between classes. Unilateral motor imagery has been shown to generally result in event-related desynchronization (ERD) in SMR at contralateral hemisphere \cite{pfurtscheller2006mu,pfurtscheller1997motor}. The informative features from sensorimotor area were closely related to ERD and carried important discriminant information for EEG decoding of motor imagery tasks \cite{Blankertz1}. We further measured the discriminability of the selected features using r-square discriminant index \cite{blankertz2011single}. The feature discriminability shown in Fig. \ref{F-SelectedFeatures} indicates that multi-task learning selects more features with higher disscriminability in comparison with single-task learning, which are conducive to better SMR-related EEG pattern decoding accuracy. Hence, these analyses provide explicit evidences for the superior performance of our proposed algorithm over the single-task learning-based approaches.

\subsection{Multiple bands versus canonical bands}
\begin{figure}[!t]
	\centering
	\includegraphics[width=0.4\textwidth]{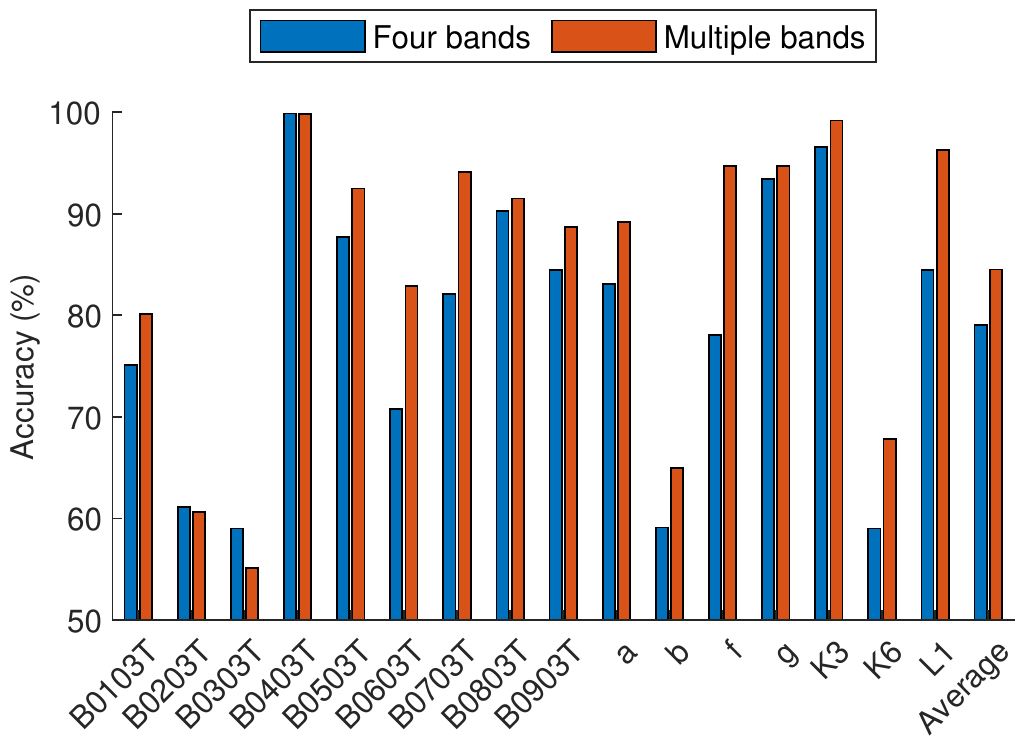}
	\caption{Comparison of EEG decoding accuracy between using multiple bands and canonical bands. For multiple bands, we used 17 overlapping subbands (as introduced in section III-B). For canonical bands, we used four frequency bands including delta (1--3 Hz), theta (4--7 Hz), alpha (8--13 Hz), and beta (14--30 Hz) that have been popularly used in various EEG study.}
	\label{F-srMTLAccComparison}
\end{figure}

In our experimental study, we extracted EEG features using multi-band optimization strategy  with multiple overlapping subbands (as introduced in section III-B). For brain signal analysis, some canonical frequency bands including delta (1--3 Hz), theta (4--7 Hz), alpha (8--13 Hz), and beta (14--30 Hz) have been popularly used in various EEG study. Here, we also investigated how the selection of filter bands would affect the performance of our proposed srMTL algorithm. EEG decoding accuracy was compared between using multiple overlapping subbands and the four canonical bands for each of the subjects across all the three independent datasets (see Fig. \ref{F-srMTLAccComparison}). The results indicated that srMTL with multiple overlapping subbands yielded much better accuracy for most of the subjects. The multiple subbands provide richer band information that allow the feature selection methods (e.g., our algorithm) to more accurately capture the subject-specific, thereby enhancing decoding accuracy, compared with a relative wide frequency band \cite{FBCSP,SFBCSP}.

\subsection{Parameter sensitivity}
\begin{figure*}[!t]
	\centering
	\includegraphics[width=0.85\textwidth]{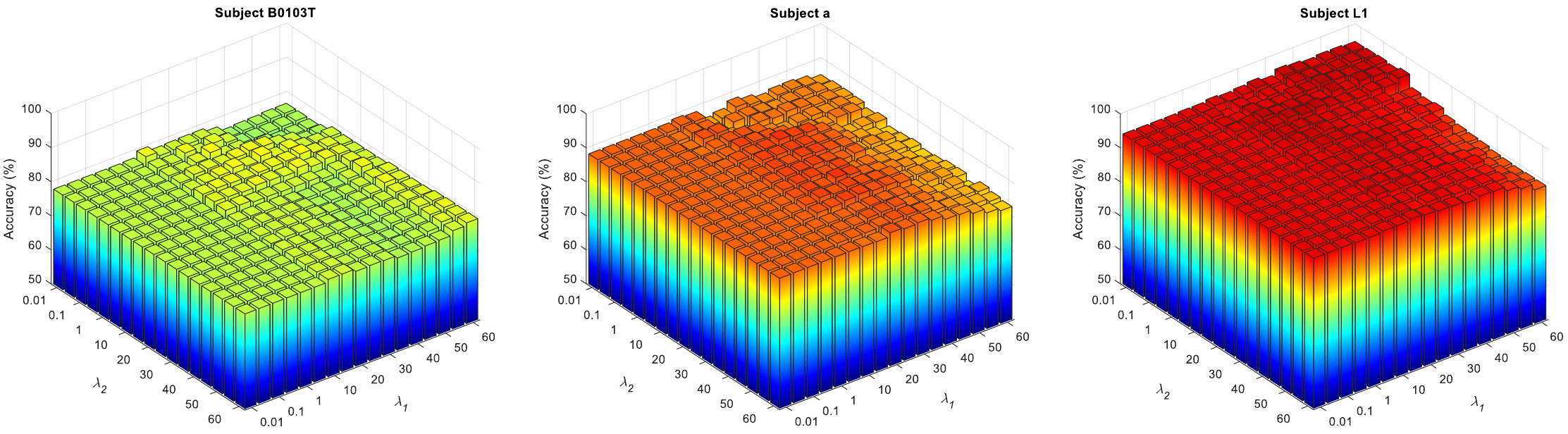}
	\caption{Parameter sensitivity of our proposed algorithm. The three figures depict the effects of varying the regularization parameters $\lambda_1$ and $\lambda_2$ on the classification accuracy of srMTL for subjects B0103T, a, and L1, respectively. These show that our algorithm achieves generally stable performance with slight fluctuation across a relatively large range of the hyperparameter values, demonstrating its robustness to hyperparameter changes.}
	\label{F-Parameter}
\end{figure*}

It should be noted that the performance of our proposed srMTL algorithm depends on the selection of hyperparameters $\lambda_1$ and $\lambda_2$, to some extent. We tested the parameter sensitivity of our method by independently changing the values of these two hyperparameters and evaluating the corresponding classification accuracy. Fig. \ref{F-Parameter} depicts the effects of varying the two hyperparameters on the classification accuracy of srMTL, for three subjects from each of the three studied datasets. The results show that the performance of srMTL slightly fluctuates with the changing parameter values across a relatively large range, demonstrating the general stability of our algorithm with respect to the hyperparameters.

\begin{figure}[!t]
	\centering
	\includegraphics[width=0.4\textwidth]{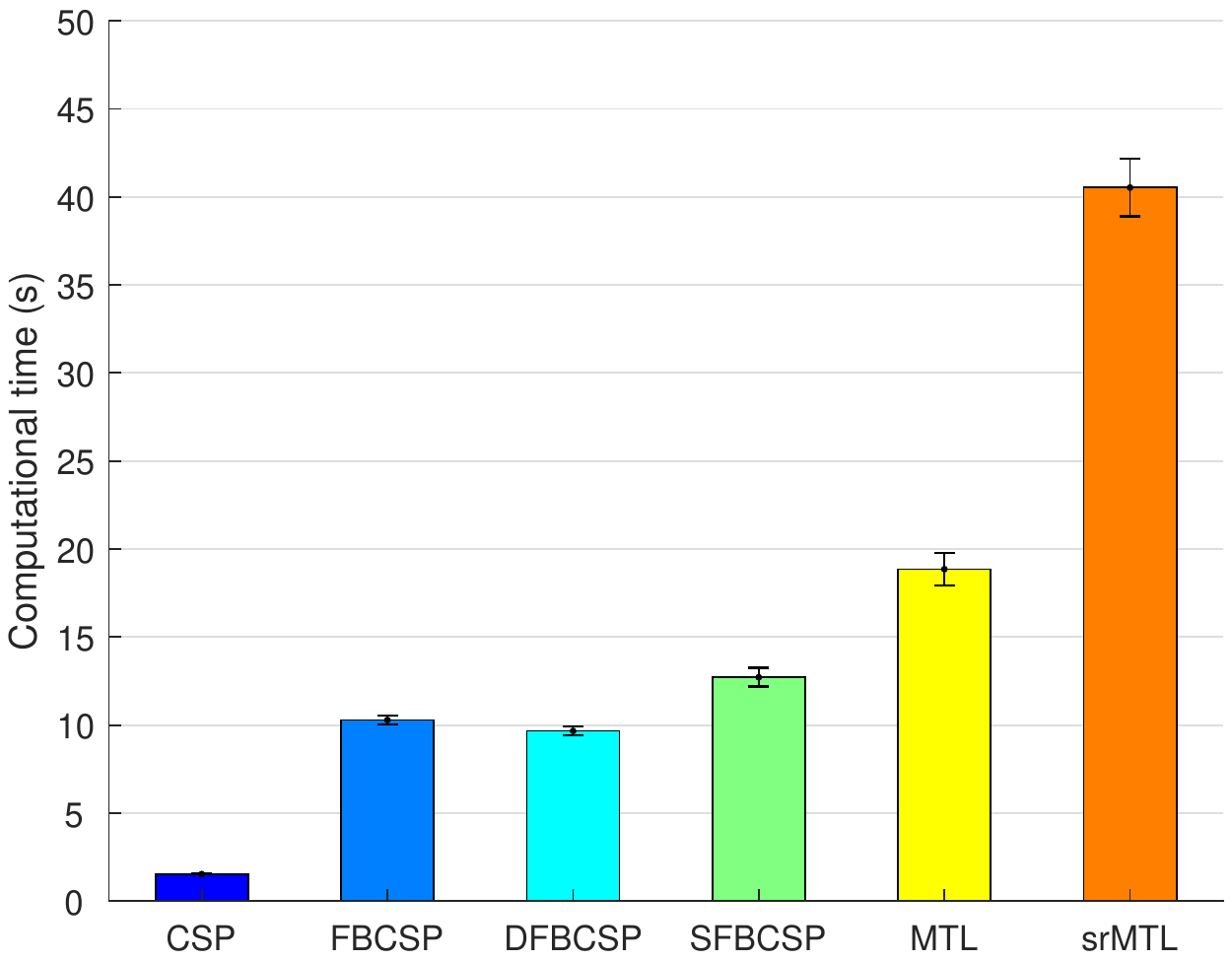}
	\caption{Comparison of computational efficiency between different algorithms for EEG classification.}
	\label{F-CompTime}
\end{figure}

\subsection{Computational cost}
To confirm the efficacy of our proposed srMTL algorithm, we also evaluated its computational efficiency in comparison with other methods (See Fig. \ref{F-CompTime}). All the methods were implemented for $5 \times 5$-fold cross-validation with MATLAB R2017a on a computer (3.41 GHz CPU, i7-6700U, 64GB RAM) and the computational time was averaged across runs of the cross-validation. Although our method took a longer time than other compared approaches, all of them achieved high computational efficiencies that meet the practical needs of BCI systems. Most of the computational cost of our algorithm arises from the inner-loop cross validation for hyperparameter selection, which will not affect the online performance of BCI. With improved high performance of computation using GPU in our future study, we expect to dramatically reduce the computational cost to speed up the training phase. More importantly, our algorithm consistently outperformed other compared methods for all of the three independent datasets with significantly enhanced EEG decoding accuracy, demonstrating its great potential in the development of improved BCI systems.

\subsection{Limitations and extension}
Our current experimental study only tested the performance of our algorithm on EEG classification for motor imagery-based BCI. It would be straightforward to extend our method to decoding other types of EEG patterns, such as event-related potential and visual evoked potential. This will be investigated in our next studies.

In our experimental study, we adopted an inner-loop cross-validation procedure to determine the appropriate hyperparameter values for the proposed algorithm and showed its robustness. However, implementing the cross-validation requires additional samples for performance validation and is generally time-consuming, which limits the practicability of BCI systems, to some extent. Bayesian inference provides an elegant way to circumvent this issue by exploiting a properly designed prior distribution \cite{wu2015probabilistic,wu2016bayesian}. As a typical method, sparse Bayesian learning-based algorithms \cite{jin2018eeg,ZhangSBC,zhang2017sparse,ZhangNHB2019} have been developed for automatic optimization of model hyperparameters, by exploiting a sparsity-induced prior, such as automatic relevance determination or Laplace distribution. Thus, the EEG decoding may further benefit from the probabilistic extension of the proposed algorithm, which is worth further study.

In the proposed algorithm, we applied AP clustering algorithm with an automatically determined number of clusters for subclass discovery in the EEG data. It is worth further investigating how our algorithm would work with other types of clustering algorithms, such as k-means and hierarchical clustering. On the other hand, although the AP clustering algorithm has been shown to work well for many clustering applications \cite{frey2007clustering,dueck2007non}, no feature selection has been taken into account for the clustering. Such a clustering procedure without feature selection may not provide the optimal solution since the true underlying subclasses present in the data may differ only with respect to a subset of the features. By exploiting a sparsity-induced constraint, sparse clustering algorithm \cite{witten2010framework} has been developed to achieve simultaneous feature selection and clustering in a data-drive way. Thus, we believe that modifying our model by incorporating sparse clustering may further improve the algorithm performance. On the other hand, multi-view clustering \cite{zhang2018binary} and discriminative embedding dictionary learning \cite{li2019discriminative} have proven to be powerful to uncover the intrinsic data structure from multi-view information. In addition, some recent studies \cite{zhang2016l2,li2015robust} has explored unsupervised clustering and $l_{2,p}$-norm regularization for feature learning, which effectively improved the classification performance. Incorporating these strategies into our designed algorithm will also be an interesting direction for future study.

In recent years, deep neural networks have attracted increasing interests and proven their strength in various pattern recognition applications \cite{nie2018strainet,zhou2020hi,liu2018exploiting,zhang1905survey,liu2019weakly,gao2019eeg,kwon2019subject,sakhavi2018learning}. For example, a deep collaborative embedding method \cite{li2018deep} has been developed for the optimal compatibility of representation learning and latent space discovery, which achieved promising performance for multiple image understanding tasks. Accordingly, we consider that extending our multi-task learning-based algorithm into a convolutional neural network or long-short term memory recurrent neural network to optimize the features with higher-level representation may further boost EEG decoding performance. This will be studied in our future work.

\section{Conclusions}
In this study, we proposed a clustering-based multi-task learning algorithm to optimize the task-related EEG feature for improved pattern recognition of brain activities in BCI applications. Specifically, we explored the potential sub-classes using affinity propagation clustering algorithm to characterize the intrinsic sample structure of EEG data. With the encoded label matrix, we further designed a subclass relationship constrained feature learning algorithm based on a multi-task learning framework to jointly optimize EEG features from the uncovered sub-classes. Followed by a linear support vector machine trained on the optimized features, our proposed algorithm provided significantly improved classification performance in comparison with other state-of-the-art algorithms. Our future studies will investigate the performance of our algorithm on other types of BCI systems.


\bibliographystyle{IEEEtran}
\bibliography{reference}

\end{document}